# FORECAST OF THE CHEMICAL AGING
# AND RELATED COLOR CHANGES IN PAINTING

B. Zilbergleyt
System Dynamics Research Foundation, Chicago
livent@ameritech.net

## ABSTRACT

The article describes the potential application of thermodynamic simulation to the problems of chemical aging of painting. Qualitative and numerical results were obtained in a preliminary investigation by applying the method to various mixtures of pigments without and with atmospheric components. The results were compared to the legendary recommendation on incompatible pigment mixtures with about an 80% match regarding potential color changes in the aged mixtures of pigments. Results for the cadmium yellow – lead white and cadmium lemon - emerald green mixtures are illustrated by pictures, gradually showing color changes related to the aging. The method of thermodynamic simulation can be a powerful tool to investigate old masterpieces, in developing new materials, and to forecast some aspects of the aging of real masterpieces.

## INTRODUCTION

Color is the first and quite often the major agent binding the painting with the viewers, and color is the first to suffer at the aging of a painting. Often starting with fading, color changes in paint layers after a certain time may heavily damage the whole picture. Old painting collections in the world museums abound with darkened canvases because of painting vulnerability to many detrimental, external and internal impacts. The age of the old tempera paintings estimated as one thousand years; usage of the oil binders is probably as twice as younger [1]. Both had enough time, mostly in unfavorable conditions, to show heavy changes really occurred in color and picture. Works of old Flemish masters were kept for a long time in their enemies' hands before being placed in the museums. At a glance, the Van Gogh's heritage had more luck; his works have been appreciated for only about 5 decades after his death. But general neglect of the good taste in painting techniques and materials, typical for the second half of 19[th] century, didn't bypass his works, and many of them showed definite results of the chemical deteriorating processes within the paint layers [2]; a story of Van Gough's "Roses" was told in [3].

Though external enemies – climatic factors (temperature, humidity, ultraviolet radiation), bacteria and reckless people as well – are always at work, their efforts can be tangibly restricted in many cases. But in the long term the major reasons for changes of the color of a painting result from the chemical aging of the paint layers. Those changes include changes in the optical properties of binders, and changes in chemical and structural composition of pigments due to chemical interactions between them and with atmospheric compounds. Chemical deterioration due to interaction between paint layer components, assisted by atmospheric species, looks to be immanent to the aging process.

The patterns of the aging of painting have been intensively studied, and seem to be quite clear on a qualitative level (results and ideas are summarized briefly in [4]). Original paintings are usually well protected against chemical impact. Being a complex composition of grounds, pigments, organic binders, and varnish, in an ideal case the fresh paint layer contains well-encapsulated pigment particles. The particles are separated from the base (canvas, wood, etc.) by grounds, from each other and the ground - by the binder, and from the atmospheric air - by both the binder and a varnish layer. As long as the grounds, binders and varnish are intact, routine changes of paint layers are extremely slow. But even in the most favorable storage conditions, the upper layer often ages faster than the inner layers. Initially it leads to darkening due to a change in the optical characteristics of the varnish, and then multiple cracks inevitably open up. From the opposite side, binders and grounds also can crack due to the temperature and humidity variations, periodical removals of the canvas from stretcher and placing it back, or due to drying of the wood basis. To make the grounds more elastic, artists used animal glue, particularly fish glue, sometimes adding honey as plasticizer – all those components are so tasteful to bacteria that glue layers and grounds easily undergo the biodegradation. Protection of the color bearing paint layer becomes more and more illusive. Moisture and some atmospheric components, penetrating through the cracks, sharply accelerate chemical interactions between pigments and other ingredients of the layer thus leading to intensive changes in its chemical and phase composition. Sooner or later, our eye can identify besides cracks the massive changes in the color, brightness and contrast of the painting. Finally it calls for restorer. The restorer tries all his/her best to bring the painting to a "like new" look, but there is no guarantee that then it looks like it was created. To the best of the author's knowledge, no one of old masters ever took a color photo picture of his masterpiece.

A painting is unique. Probably, chemical and relevant color changes at the first stages of aging may turn it to a more unique thing, but anyway it will be different from the original. What can we do and to what extent can we prevent the chemical changes? In contrast to bio and physical reasons of aging, chemical aging – as we understand it - has been never deeply studied. We know some very general, legendary old masters' recommendations on the incompatibility of certain couples of inorganic pigments, like "white lead – ultramarine" and some others, originated from long time experience [5]. Besides that the artist also has to take into consideration certain technological approaches. In any case, we need a quantitative picture of what's going on at the aging.

## THE METHOD OF THERMODYNAMIC SIMULATION

The method of thermodynamic simulation consists of the computer simulation of possible chemical interactions within the paint layer to determine the ultimate chemical and phase equilibrium composition. It allows for numerical calculations of the most probable final composition of the paint layer, resulting from these interactions when all changes are over and the system rests at thermodynamic equilibrium. One can state that in old art pieces most of the possible chemical processes, allowed by their pre-museum and museum storage conditions, are either completed, or the relevant changes are already well pronounced. Knowledge of the chemical and phase changes in the layer may help to predict optical/color changes and such structural damages (like the paint layer dropping down).

The method can be implemented using most of the known simulation software; we used the simulation complex ASTRA-4 [6] and partly HSC Chemistry [7]. A pressure of 0.1 Pa and a temperature of 293K ($20^0$C) were taken as simulation parameters. In case of pair mixtures of pigments, the mass ratio of main to admixed components varied from 10:1 to 1:1. Taking into account very long life time of the painting, atmospheric air as a natural mixture of oxygen and nitrogen was presented in the initial compositions up to 10%, with moisture up to 5% of the air mass and some typical pollutants like CO, $CO_2$, sometimes $SO_2$ and $H_2S$ up to concentrations, typical for big cities.

In cases when not enough thermodynamic data was available to involve some complex pigments of interest into simulation, we used their essential and often the major color carrying fragments. For example, in case of lead white we used the data for lead carbonate only; in this case simulation was carried out with the fragment of the pigment with known data instead of the whole compound. The same situation occurred in case of ultramarine and several others. Though in most cases this was good enough to judge the possible color changes, the results of this work should be considered qualitative for manifesting the method's ability to predict and describe the chemical aging of painting.

## THE SIMULATION RESULTS

We tried chemical interactions within several groups of pigments and atmospheric components, including interactions of various pigments with atmospheric components; white lead and zinc white with some pigments in presence of atmospheric components; oxide pigments in mixtures with sulfide pigments; special pairs of pigments to check the legendary incompatible pigment couples; some complex mixtures from the P. P. Rubens palette.

Among them the group of "tabooed" incompatible couples of pigments is one of the most interesting in the light of this work. Corresponding results are shown in Table I. Empirical legendary results were taken from [4]. In some cases, when dark and white substances were formed in the mixture together (such as **FeS**, CaS, and Na$_2$CO$_3$ in ultramarine mixtures with ochers and umbers), it was difficult to

Table I.

### COMPATIBILITY OF VARIOUS PIGMENTS IN PAIR MIXTURES

N – not compatible, Y – compatible, Emp. – empirical, Sim. - simulated. Black or just dark colored species, arrived in result of chemical interactions, are typed in bold.

| Pigments | | Empirical optical changes | Simulated | | Compatibility | |
|---|---|---|---|---|---|---|
| Base | Admixed | | New species able to change the colors | Potential color changes | Emp. | Sim. |
| White lead | Ultramarine | Darkening or browning | Na$_2$CO$_3$, **PbS**, **PbO$_2$** | Browning | N | N |
| | Caput-Mortuum | | | | N | N |
| | English Red | | | | N | N |
| | Cobalt Blue | | none | | N | Y |
| | Dark Ochre | Tone brightening | **Pb** | Darkening | N | N |
| | Natural Umber | | **Pb, PbO** | | N | N |
| | Cinnabar | None | None | None | Y | Y |
| | Copper pigments | None | None | None | Y | Y |
| Cadmium Yellow | White Lead | Blackening | **Pb**, CdSO$_3$ | Blackening | N | N |
| | Massikot | | **Pb** | | N | N |
| | Lead Crone | | **Pb, PbO**, Cr$_2$O$_3$ | | N | N |
| | Ultramarine | Brightening | None | None | N | Y |
| | Iron Oxides | Browning | **Fe, Fe$_3$O$_4$**, CdSO$_3$ | Darkening | N | N |
| | Umbers | | **Cd, MnS**, CaS | | N | N |
| | Terra di Verona | None | **Cd**, CdSO$_3$, MgO | Probably none | Y | Y |
| Cobalt pigments | Ultramarine | Changing tone | **Co**, Na$_2$SO$_4$ | Darkening | N | N |
| | Mn-Cd paint | ? | **Co** | | N | N |
| | Cadmium Yellow | None | **Co** | | Y | N |
| Ultra-marine | Ochres | Tone brightening and changing | **FeS**, CaS, Na$_2$CO$_3$ | Brightening | N | N |
| | Umber | | | | N | N |
| | Copper pigments | "Some changes" (?) | **Cu$_2$S**, NaOH | Darkening | N | N |
| Cadmium orange/red | Umbers | None | **MnO** | Probably graying | Y | N |

evaluate if optical changes would happen. Also, gaseous reaction products are not shown in the table. Even with this restriction, in about 80% of investigated couples the predicted possible optical changes have certainly matched the legendary changes. What is remarkable is that the method of thermodynamic simulation also explains possible reasons of the changes in terms of chemical and structure composition thus allowing for visual interpretation. Percentage of the matching Y/N compatibility results in Table I is high enough to prove the applicability of the method to the analysis of chemical aging of painting at least on the qualitative level. The quantitative results definitely will bring up more information though they are harder to get. Table II contains numeric simulation results of the incompatible couple mixture of lead white (represented by lead carbonate) with yellow cadmium (represented by cadmium and zinc sulfides). The abbreviations for colors are **B** for black, W for white or of light colors, and Y for yellow. To take into account the lead carbonate binding in $2PbCO_3*Pb(OH)_3$, its thermodynamic activity coefficient was reduced to 0.6 for simulation. One can see the changes in the chemical composition of the mixture due to decay of input basic color carriers, sure structural changes due to arrival of new components, and drastic color changes with arrival of black lead sulfide. Fig.1 shows color visualization of the simulation results, RGB indices were calculated using data for yellow cadmium from [8]. Colors of mixtures were calculated as weighted mean of the ingredients' colors [9].

Table II.

NUMERICAL SIMULATION RESULTS FOR
THE YELLOW CADMIUM – LEAD WHITE MIXTURE CHEMICAL AGING.

| Species | PbCO3 | CdS | ZnS | **PbS** | ZnCO3 | CdCO3 | R | G | B |
|---|---|---|---|---|---|---|---|---|---|
| Initial amount, mol. | 1.000 | 0.500 | 0.500 | | | | 256 | 185 | 137 |
| Equilibrium amount, mol | 0.538 | 0.378 | 0.160 | **0.462** | 0.339 | 0.122 | 159 | 139 | 113 |
| Color | W | Y | | **B** | W | W | | | |

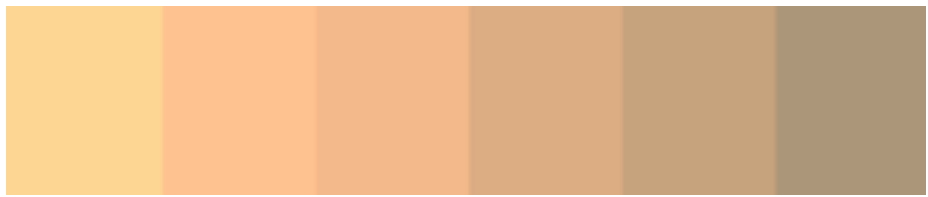

Fig.1. Color changes following chemical aging of 1:1 mixture of yellow cadmium with lead white, consequent stages at equal time intervals. The leftmost sample corresponds to a fresh mixture, the rightmost – to the mixture, aged to the equilibrium limit.

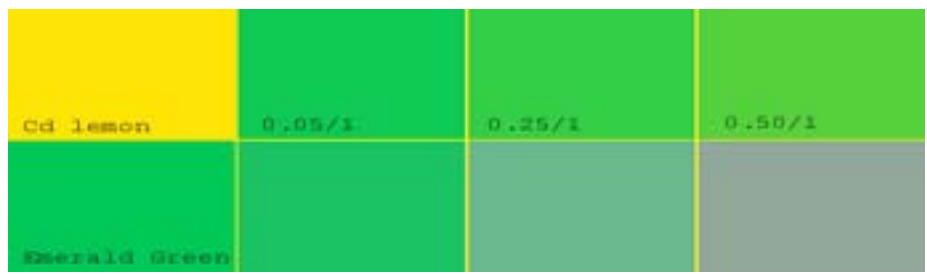

Fig.2. Color changes following chemical aging of the lemon cadmium mixtures with emerald green, various initial mixture contents. The leftmost swatches show accepted colors of pure mixture components.

Fig.2 shows the aged samples of the cadmium lemon mixture with emerald green, compared to the initial colors (upper row); numbers on the swatches show initial moles ratio Cd/Emerald Green in the mixture. Interestingly, that Cd-pigment almost totally disappears in all investigated mixtures after ageing; its increase in the initial mixture leads to more and more pronounced grey color due to formation of dark, fine dispersed particles of $Cu_3As$ and $Cu_2O$ along with several white products.

## CHEMICAL AGING OF SOME RUBENS' PIGMENT MIXTURES

Rubens was one of those rare artists who thoroughly wrote down the major components of the mixtures he used in various purposes, sometimes with a kind of functional names [1], that allows us to experiment with his palette. Qualitative results of the simulation of the mixtures aging in presence of normal air (with typical some moisture) are placed in Table III.

Table III.

PREDICTED CHANGES IN VARIOUS PAINT MIXTURES OF P. P. RUBENS' PALETTE.

| Color of the mixture | Pigments | | Initial mixture for simulation | Simulation results | |
|---|---|---|---|---|---|
| | base | admixture | | New species after aging | Possible changes |
| Neutral white | Lead white | Cinnabar, ochre, organic black | Lead white, cinnabar, ochre, air | **Hg, PbO₂, H₂SO₄** | Darkening |
| Body color | Lead white | Cinnabar, krapp-laquer, ochre, lapis lazuli, azurite, organic black | Lead white, ochre, ultramarine, azurite, air | **Hg, PbS, Cu₂S**, Na₂SO₄, Na₂CO₃ | ? |
| Blue | Lapis lazuli, indigo | Azurite, cinnabar, ochre, lead white, organic black, smalt | Ultramarine, ochre, cinnabar, azurite, lead white, air | **PbS, Hg, Cu₂S**, Na₂SO₄, Na₂CO₃, NaOH | ? |
| Yellow/ Brown | Ochre | Lead white, organic black, krapp-laquer, lapis lazuli, smalt, azurit | Ochre, lead white, azurite, ultramarine, air | **Fe, FeS, Pb, Cu**, Na₂CO₃ | Darkening |
| Green | Azurite, malachite | Lead white | Azurite, white lead, air | **CuO** | Darkening |

## CONCLUSION

This work presents the results of computer experiment in the field where all knowledge was and still is very approximate, collected from judgments of old paintings with not precisely known initial compositions. The data, related to paintings aging and pigments compatibility nomads from one printed or electronic source to another, quite often without clear references. In spite of importance of this knowledge, no one can indicate any solid collection or a survey. Certain information definitely can be achieved after centuries of waiting or using accelerated methods, usually run at conditions different than natural (like elevated temperatures). In most cases the verbal level of description of the changes is prevailing in the art aging research up to present time ([10], a good example). To the best of the author's knowledge, our approach is the first attempt to use an experimental method to replace word-of-the-mouth legends.

Discussing the results of this work, one should keep in mind that thermodynamic simulation brings up the ultimate results of aging – potential chemical and structural changes providing that all possible interactions in the ground-pigment-binder-varnish system are over, that is in thermodynamic equilibrium. Equilibrium compositions, in their turn, should be thought as the limit, to which the system tends to move, but not necessarily achieves it in reality. Chemical aging runs to a full gear at the latest stages of the life of a painting, and is just one of possible major contributors to the aging processes.

Though the scope of this investigation was restricted to inorganic pigments, the offered method is applicable to analyze chemical interactions between any kind of substances including organic dyes, binders and grounds. Gigantic color and computational abilities of the current computers essentially expand opportunities for simulation and output of the simulation results in such complicated objects like painting mixtures. Besides that, the method can be used to investigate chemical behavior not only of a painting but also of any art materials that change by interacting with their environment. Although the method investigates only one important aspect of the aging of paintings, in certain cases its results can explain and help the restorers better than other approaches.

It is not very simple to use the method of thermodynamic simulation in the field of aging because only rare pigments are chemically simple enough to have their thermodynamic properties ready in the regular thermodynamic data bases. Calculation or experimental finding appropriate information would be the first task to be solved prior to implementation of the method.

The author foresees many objections related to the results of this article. Among them – why most of the species, predicted in the run of that preliminary investigation, were never reported by previous investigators? Well, it could be that the art pieces, investigated earlier on, didn't achieve the extreme stages that could be qualified as equilibrium in the context of this article. Or – what if they were just overlooked because nobody suspected them to be present?

Who could be interested in using this method? It can definitely help to investigate possible reasons for the changes that happened, and the changes that will happen. It could be useful for attribution purposes, solving the so called back task of simulation - to find out the initial composition of the paint mixture, having the chemical analysis of the aged sample. The manufacturers of art materials could also use the method to evaluate their products in various mixtures and environmental conditions, for instance, to find proper temperatures for the expedited aging research. As concerns to the artists, it's of a very low probability that Jackson Pollock would ever be willing to hear about any simulation program, but P. P. Rubens definitely might be interested to use it.

## ACKNOWLEDGEMENT

The original preprint of this article, prepared in the main frame computers era, was published in a small amount of copies by the Ministry of Culture of the USSR [11] with support of Dr. L. Gorel'chenkova from the Moscow Institute of Art Restoration. The author also obliged to Dr. E. Carter from the "Color" Journal for a certain inspiration to proceed with this publication. It is my pleasure to express sincere gratitude to my long time editor, Dr. C. Nolte from the EditAvenue.com for his always very useful help.